\documentstyle[prl,aps,epsf]{revtex} 

\begin{document}

\title{Angular Dependence of the Upper Critical Field \\
in Triplet Quasi-One-Dimensional Superconductors
at High Magnetic Fields}

\author{Cawley D. Vaccarella and C. A. R. S\'a de Melo} 
\address{School of Physics, 
Georgia Institute of Technology, Atlanta, GA 30332} 

\date{\today}
\maketitle

\begin{abstract}
Assuming that the order parameter corresponds to 
an equal spin triplet pairing symmetry state, 
we calculate the angular dependence of the 
upper critical field in triplet quasi-one-dimensional superconductors 
at high magnetic fields applied along the $yz$ plane. 
\vspace{1pc}
\end{abstract}


The upper critical field of quasi-one-dimensional superconductors 
for perfectly aligned
magnetic fields along the $y$ $(b^{\prime})$ axis 
has been calculated for both singlet and triplet 
states~\cite{lebed-86,lebed-87,dms-93}. 
More recently new experiments performed in these systems
lead to the observation of unusual superconductivity 
at high magnetic fields~\cite{lee-95,lee-97}
and created new excitement in the area of organic 
superconductivity~\cite{shimahara-97,jerome-97,hasegawa-99,ishiguro-00}.
In quasi-one-dimensional superconductors (of the 
Bechgaard salts family with chemical formula ${\rm (TMTSF)_2 X}$, 
where ${\rm X = ClO_4, PF_6, ...}$) the upper critical field exceeds 
substantially the Pauli paramagnetic
limit~\cite{lee-95,lee-97}. 
This indicates that high magnetic field superconductivity 
in these systems is most likely triplet. The exact symmetry of the 
triplet phase and the mechanism of superconductivity are not yet known, 
however an early suggestion of triplet superconductivity (at zero 
magnetic field) was given by Abrikosov~\cite{abrikosov-83} in order
to explain the suppression of 
superconductivity in ${\rm (TMTSF)_2 X}$ in the 
presence of disorder 
(non-magnetic impurities)~\cite{choi-82,green-82}.

Very recently, it has been argued 
by Lebed, Machida, and Ozaki~\cite{lebed-00} (LMO)
that the spin-orbit coupling must be strong in order to explain 
the observed experimental upper critical fields and the absence of the 
Knight shift for fields parallel to the $y$ $(b^{\prime})$ 
axis.~\cite{lee-00}
However, we find this possibility hard to be realized in the
${\rm (TMTSF)_2 X}$ family
since any reasonable estimate of the atomic spin-orbit coupling 
leads to a very small value (in addition,
the Bechgaard salts are very low disorder systems, 
with very long mean free paths).
Estimates of the value of spin-orbit coupling 
are several orders of magnitude smaller~\cite{lee-98} 
than the values required to fit the critical
temperature of ${\rm (TMTSF)_2 PF_6}$~\cite{lebed-00,maki-89}, 
for instance, at low magnetic fields.
That the atomic spin-orbit coupling should 
be small is not surprising, since the
heaviest element in the ${\rm (TMTSF)_2 X}$ family is ${\rm Se}$.
Thus, unlike in the work of LMO, we consider here only the weak 
spin-orbit coupling limit corresponding to the equal spin triplet
state (ESTP) proposed by one of us~\cite{sdm-99,sdm-96,sdm-98}.
Here we also take into consideration the effects of time 
reversal symmetry breaking into the structure of the order parameter,
for finite magnetic fields. We would like to mention, in passing,
that the ESTP state is also consistent with the absence of the
Knight shift for fields parallel to the $y$ $(b^{\prime})$~\cite{lee-00},
predicting in fact that the low field susceptibility in 
the superconducting state at zero temperature 
$\chi_s (T = 0) = \chi_n$, the susceptibility at the normal state for 
any direction of the applied field~\cite{sdm-99,sdm-96,sdm-98,chi-00}. 

We focus in this paper only on the angular dependence
of the upper critical field, when the magnetic field is applied 
along the $yz$ plane and show that deviations
from perfect alignment along the $y$ $(b^{\prime})$ axis
cause a dramatic drop in the critical temperature for 
intermediate values of the magnetic field in the range of $5$ to $30$
Teslas. We then argue that this rapid suppression of the critical 
temperature is in qualitative agreement with the rapid drop of the
putative upper critical in ${\rm (TMTSF)_2 ClO_4}$~\cite{lee-98}.


We model quasi-one-dimensional systems 
via the energy relation
\begin{equation}
\label{eqn:dispersion-spin}
\varepsilon_{\alpha, \sigma}({\bf k}) =
\varepsilon_{\alpha} ({\bf k}) - \sigma \mu_B H,
\end{equation}
with the $\alpha$-branch dispersion 
\begin{equation}
\label{eqn:dispersion-0}
\varepsilon_{\alpha} ({\bf k}) = v_F (\alpha k_x - k_F) + 
t_y \cos(k_y b) + t_z \cos(k_z c),
\end{equation}
corresponding to an orthorombic crystal with lattice constants
$a,b$ and $c$~\cite{triclinic-00} along the $x,y$ and $z$ axis 
respectively. 
In addition, since $E_F = v_F k_F/2 \gg  t_y  \gg t_z $,
the Fermi surface of such systems is not simply connected, 
being open both in 
the {\it xz} plane and in the {\it xy} plane. Furthermore, the 
electronic motion can be classified as {\it right}-going ($\alpha = +$) 
or {\it left}-going ($\alpha = -$). 

We begin our discussion by analysing the electronic motion 
for magnetic fields ${\bf H } = (0, H_y, H_z)$ in $yz$ plane.
We work in units where the Planck's constant and the 
speed of light are equal to one, i.e., $\hbar = c^* = 1$, 
and, throughout the paper, we use the gauge 
${\bf A} = (H_y z - H_z y, 0, 0)$, where $\alpha$ and $k_x$ are
still conserved quantities (good quantum numbers)
while $k_y$ and $k_z$ are not.
In this gauge, the 
semiclassical equations of motion become
\begin{equation}
\label{eqn:xt}
x = x_0 + \alpha v_F t,
\end{equation}
\begin{equation}
\label{eqn:yt}
y = y_0 + \alpha (t_y b/ \omega_{c_y}) 
\cos( k_y(0)b + \alpha \omega_{c_y}t ),
\end{equation}
\begin{equation}
\label{eqn:zt}
z = z_0 - \alpha (t_z c/ \omega_{c_z}) 
\cos( k_z(0)c - \alpha \omega_{c_z}t ),
\end{equation}
with $\omega_{c_y} = v_F G_y$ and $\omega_{c_z} = v_F G_z$, 
while $G_y = |e|H_z b$ and $G_z = |e| H_y c$, where
$H_z = H \cos\theta$ and $H_y = H \sin\theta$, where $\theta$ is the
angle between ${\bf H}$ and ${\bf \hat z}$. 
The frequencies $\omega_{c_y}$ and $\omega_{c_z}$ are independent 
and characterize the periodic motion along the {\it y} direction 
and {\it z} direction, respectively. 

This semiclassical analysis suggests a classification of three different
regimes of confinement for the electronic motion: (a) a one-dimensional 
regime, where the semiclassical amplitudes of motion are confined along
both the $y$ and $z$ direction, i.e., 
$t_z/\omega_{c_z} \ll 1$ or $t_y/\omega_{c_y} \ll 1$ (provided that there
is no magnetic breakdown, and that $E_F \gg \omega_{c_y}$ and $E_F \gg 
\omega_{c_z}$); (b) a two-dimensional regime, where 
the semiclassical amplitudes of motion are confined along
the $z$ direction but not along the $y$ direction, i.e., 
$t_z/\omega_{c_z} \ll 1$ or $t_y/\omega_{c_y} \gg 1$; and 
(c) a three-dimensional regime, where the semiclassical amplitudes of motion
are not confined in either $y$ or $z$ directions, i.e.,
$t_y/\omega_{c_y} \gg 1$ or $t_z/\omega_{c_z} \gg 1$. 
The confinement (localization) first occurs along the
{\it z} direction when 
$|t_z c/ \omega_{c_z}| \leq |t_y b / \omega_{c_y}|$
which implies $|\tan\theta| \ge t_z/t_y$, i.e.,
when $\theta \geq \theta_c = \arctan (t_z/t_y)$.
The localization first
occurs along the {\it y} direction when $|\tan\theta| \leq t_z/t_y$
or $\theta \le \theta_c$.
Hence in the limit where $t_z/t_y \ll 1$ the localization happens first 
along the {\it z} direction except for a very narrow region of angles
$\theta \leq \theta_c$. For instance,  
taking $t_z/t_y \approx 0.1$, $\theta_c \approx 5.71^{\circ}$ and the 
angular region where localization takes place first along the {\it z}
direction is very wide ($\theta_c < \theta < 90^{\circ}$), 
except for a few degrees from the {\it z} axis.

Now we turn our attention to the quantum aspects of the problem. 
In the presence of the magnetic field $H$, the non-interacting 
Hamiltonian is
\begin{equation}
H_0 ({\bf k} - e {\bf A}) = 
\varepsilon_{\alpha} ({\bf k} - e {\bf A}) - \sigma \mu_B H
\end{equation}
in the gauge ${\bf A} = (H_y z - H_z y, 0, 0)$.
The eigenfunctions of $H_0 ({\bf k} - e {\bf A})$ are 
\begin{equation}
\Phi_{qn}({\bf r}) = \exp[ik_{x}x]
J_{N_y-n_y}\left({\alpha t_y \over \omega_{c_y} }\right)
J_{N_z-n_z}\left({\alpha t_z \over \omega_{c_z} }\right), 
\end{equation}
where ${\bf r} = (x, y, z)$,  
$y = n_y b$ and $z = n_z c$ with associated quantum numbers 
$qn = \alpha,k_x,N_y,N_z,\sigma$ and 
eigenvalues
\begin{equation}
\label{eqn:eigenqn}
\epsilon_{qn} = 
\varepsilon_{\alpha, \sigma} ({\bf k}_{\rho}) + \alpha N_y \omega_{c_y}
+ \alpha N_z \omega_{c_z}.
\end{equation}
The function $J_{p}(u)$ 
is the Bessel function of integer order {\it p} and 
argument {\it u}, while now
\begin{equation}
\varepsilon_{\alpha, \sigma} ({\bf k}_{\rho}) = v_F(\alpha k_x - k_F) 
-\sigma \mu_B H 
\end{equation}
is a 1D dispersion. 
Notice that the eigenspectrum in Eq.~(\ref{eqn:eigenqn}) involves
many magnetic subbands labeled by the quantum numbers $N_y$ and $N_z$
and that the
eigenvalue $\epsilon_{qn}$ is invariant under the 
quantum number transformation
$(\alpha,k_x, N_y, N_z, \sigma) \to (-\alpha,-k_x, -N_y, -N_z,\sigma)$,
for the same spin state $\sigma$.  
In addition, these magnetic subbands are spin-split into 
{\it spin-up} and {\it spin-down} bands. Thus, Cooper pairs can be 
easily formed in a ESTP pairing state, involving electrons 
with quantum numbers $(\alpha,k_x, N_y, N_z, \sigma)$ and 
$(-\alpha,-k_x, -N_y, -N_z, \sigma)$, provided that the pairing 
interaction conserves spin (which seems to be the case for the
${\rm (TMTSF)_2 X}$ family, except for very small spin-orbit and dipolar 
couplings)~\cite{dipolar-00}.
Thus, the order parameter vector 
\begin{equation}
\label{eqn:delta-vector}
{\vec \Delta } ({\bf r}) = \left[ \Delta_{+1} ({\bf r}), 0,
\Delta_{-1} ({\bf r}) \right] 
\end{equation}
has components in the $m_s = +1$ $(\mu = \uparrow\uparrow)$ and 
$m_s = -1$ $(\mu = \downarrow\downarrow)$ channels, only. 

In the case of ESTP, the gap equation 
can be written as 
\begin{equation}
\label{eqn:gap-lawrence}
\Delta_{\mu} ({\bf r})  = \lambda_{\mu}
\sum_{N_y, N_z}
A_{N_y, N_z}^{(\mu)} 
\Delta_{\mu} ({\bf r} + {\bf S}_{N_y N_z})
\end{equation}
where ${\bf r} = (x, y, z)$, 
${\bf S}_{N_y N_z} = (0, N_yb, N_zc)$.
The matrix operator
\begin{equation}
\label{eqn:ann}
A_{N_y, N_z}^{(\mu)}  =   E_{N_y N_z} (x)
F_{N_y N_z}^{(\mu)} ( {\hat q}_x - {K}_{N_y N_z} )
\end{equation}
with ${K}_{N_y N_z} =  2 N_y G_y + 2 N_z G_z$ 
involves two terms, the prefactor
\begin{equation}
\label{eqn:enn}
E_{N_y N_z} (x) = \exp \left[- i({K}_{N_y N_z}x) \right]  
\end{equation}
and the operator function 
\begin{equation}
\label{eqn:fnn}
 F_{{N_y}{N_z}}^{(\mu)}({\hat q}_x) = \sum_{\alpha, M_{i\nu}}
\Lambda_{\mu}^{\alpha {\bar \alpha}} 
({\hat q}_x + Q_{M_{i\nu}})
W_{N_y N_z}^{\alpha {\bar \alpha}} (M_{i\nu}), 
\end{equation}
with $Q_{M{i\nu}} = (M_{1y} + M_{2y}) G_y + (M_{1z} + M_{2z}) G_z $,
and $\hat{q}_x = -i\partial/\partial x$. The weight function
\begin{equation}
\label{eqn:wnn}
W_{N_y N_z}^{\alpha {\bar \alpha}} (M_{i\nu}) = 
\Gamma_{N_y,N_z}^{\alpha} (M_{2\nu})
\Gamma_{N_y,N_z}^{\bar\alpha} (M_{1\nu}),
\end{equation}
contains the single-particle renormalization factor 
\begin{equation}
\label{eqn:gamma}
\Gamma_{N_y,N_z}^{\alpha}(M_{2\nu}) = 
P_{N_y}^{\alpha} (M_{2y}) 
P_{N_z}^{\alpha} (M_{2z}) 
\end{equation}
with coefficient
\begin{equation}
P_{N_{\nu}}^{\alpha}(M_{2\nu}) =
J_{ M_{2\nu} }
\left( { \alpha t_{\nu}\over \omega_{c_{\nu}} } \right)
J_{ M_{2\nu}- N_{\nu} } 
\left( { \alpha t_{\nu} \over \omega_{c_{\nu}} } \right). 
\end{equation}
Notice that the second term 
in the right hand side of Eq.~(\ref{eqn:wnn}), i.e.,
$\Gamma_{N_y,N_z}^{\bar\alpha} (M_{1\nu})$ can be obtained from 
Eq.~(\ref{eqn:gamma}) via the the following substitution $\alpha \to 
\bar \alpha$ and $M_{2\nu} \to M_{1\nu}$. In addition,
the operator $F_{{N_y}{N_z}}({\hat Q}_x)$ contains the Cooper singularity
contribution
$$
\Lambda_{\mu}^{\alpha {\bar \alpha}} 
(Q_x) = 
{\cal N}_{\mu} \left[ \psi \left({1 \over 2} \right) - 
{\cal R} \psi \left( { 1 \over 2} 
+ {\alpha v_F Q_x \over 4\pi i T} \right)  + 
\ln \left({\bar\Omega} \right) \right]
$$
with ${\cal N}_{\mu}$ being the spin-dependent density of states, 
and  ${\bar\Omega} = ({2 \omega_d \gamma / \pi T})$.

The physical interpretation of Eq.~(\ref{eqn:gap-lawrence}) 
is as follows. 
The order parameter $\Delta_\mu (x, y, z)$ at a given position in space
couples with the order parameter at $\Delta_\mu (x, y + N_yb, z + N_zc)$
at a different position, thus indicating that Josephson coupling 
between chains oriented along the $x$ direction occurs.
The strength of this coupling is controlled by the matrix operator 
$A_{N_y, N_z}^{(\mu)}$, which is dependent both on the magnitude of 
the magnetic field and its direction, through $G_y$ and $G_z$.
This result is extremely interesting because it shows that
the quantization effects of the magnetic field can make 
quasi-one-dimensional superconductors evolve
from a strongly coupled Ginzburg-Landau 
local regime (low magnetic fields) into 
a {\it variable range} Josephson coupled non-local regime 
(high magnetic fields).

We will now focus our attention to three limits,
where the critical temperature as a function
of the applied magnetic (or conversely the the upper critical field
as a function of temperature) can be calculated analytically.  
We will assume that the interaction $\lambda_{\mu} = \lambda$ and the 
density of states ${\cal N}_\mu = {\cal N}$ are independent of the
spin channel~\cite{lambdamu}, which implies that $\Delta_{\mu} = \Delta$ 
in the analysis that follows.  

 
{\it One-dimensional regime}: As seen in the semiclassical 
analysis of Eqs.~(\ref{eqn:xt}),~(\ref{eqn:yt}) and (\ref{eqn:zt}), 
there is double confinement of the orbital motion when 
$t_y/\omega_{c_y} \ll 1$ and $t_z/ \omega_{c_z} \ll 1$. 
Using the fact that $q_y$ and $q_z$ are good quantum numbers
one can write 
\begin{equation}
\label{eqn:deltaxyz}
\Delta (x,y,z) = \exp [i(q_y y + q_z z)] u(x),
\end{equation}
which substituted in Eq.~(\ref{eqn:gap-lawrence}) transforms it
into the generalized (two-term) Hill equation
\begin{equation}
\label{eqn:hill}
\left[
-\beta_{1d}{\ell}_x^2 {d^2 \over dx^2} +
A_{1d} - \sum_{\nu} B_\nu \cos\phi_\nu (x)
\right] u(x) = 0, 
\end{equation}
to leading non-trivial order in ${\hat q}_x$.
Here $\phi_\nu (x)  = (q_\nu s_\nu - 2G_{\nu}x)$, $\nu = y, z$, 
$s_y = b$ and $s_z = c$.
The coefficient of the first term is  
$\beta_{1d} = (1 - {t_y^2 / \omega_{c_y}^2} - {t_z^2 / \omega_{c_z}^2})$.
This coefficient renormalizes the characteristic length scale along the 
$x$ direction
${\ell}_x = \sqrt{C} v_F/T$, 
which contains the constant 
$C = 7 \zeta (3)/16 \pi^2$. 
The second coefficient is 
$$
A_{1d} = 
\left[
\beta_{1d}
\ln \left( {\bar\Omega} \right) 
+ \sum_{\nu} 
\left({ t_\nu \over \omega_{c_\nu}} \right)^2 
\ln \left\vert{ 2 \omega_d \over
\omega_{c_\nu} }\right\vert 
\right] - {1 \over \lambda {\cal N} },
$$
while 
$B_{\nu} = 
\left( { t_{\nu} / \omega_{c_\nu} } \right)^2
\ln \left\vert {\gamma \omega_{c_\nu} / 2 \pi T} \right\vert$
are the last coefficients corresponding to the 
amplitude of co-sinusoidal spatial oscillations.
Physically, Eq.~(\ref{eqn:hill}) 
corresponds to a limit of weakly coupled Josephson chains,
with magnetic field dependent Josephson couplings $B_{\nu}$. 
The highest critical temperature in high magnetic fields, 
to lowest order in 
$t_{\nu}^2/\omega_{c_\nu}^2$, occurs when 
${q_y} = {q_z} = 0$. Thus, $T_c$ is determined by the condition 
$A_{1d} = 0$, leading to 
\begin{equation}
\label{eqn:tcyz}
T_c = T_{c_{1d}} 
\left[ 
1 - \sum_\nu \left( { t_\nu \over \omega_{c_\nu} } \right)^2
\ln 
\left\vert
{\gamma \omega_{c_\nu} \over \pi T_{c_{1d}}  }  
\right\vert
\right],
\end{equation}
with prefactor
$
T_{c_{1d}}  = 
(2 \omega_d \gamma/ \pi) \exp (- 1/ \lambda {\cal N})
$~\cite{dos},
since both the amplitude $B_{\nu}$ and the period $\pi/G_{\nu}$
of $\cos (2G_{\nu} x)$ are very small.
When $t_y/\omega_{c_y} \ll 1$ and $t_z/ \omega_{c_z} \ll 1$ there is
a magnetic field induced double quantum confinement (localization) 
effect along both the $y$ and $z$ direction, 
for almost all $\theta$. However, localization along the 
$y$ direction is harder to achieve in the Bechgaard salts 
${\rm (TMTSF)_2 X}$
${\rm (X = ClO_4, PF_6, ...)}$, since $t_y$ is typically of the order 
of a $100 {\rm K}$, thus requiring magnetic fields 
the order of $100{\rm T}$ to take place. 

\begin{figure}
\begin{center}
\epsfxsize=8cm
\epsfysize=8cm
\epsffile{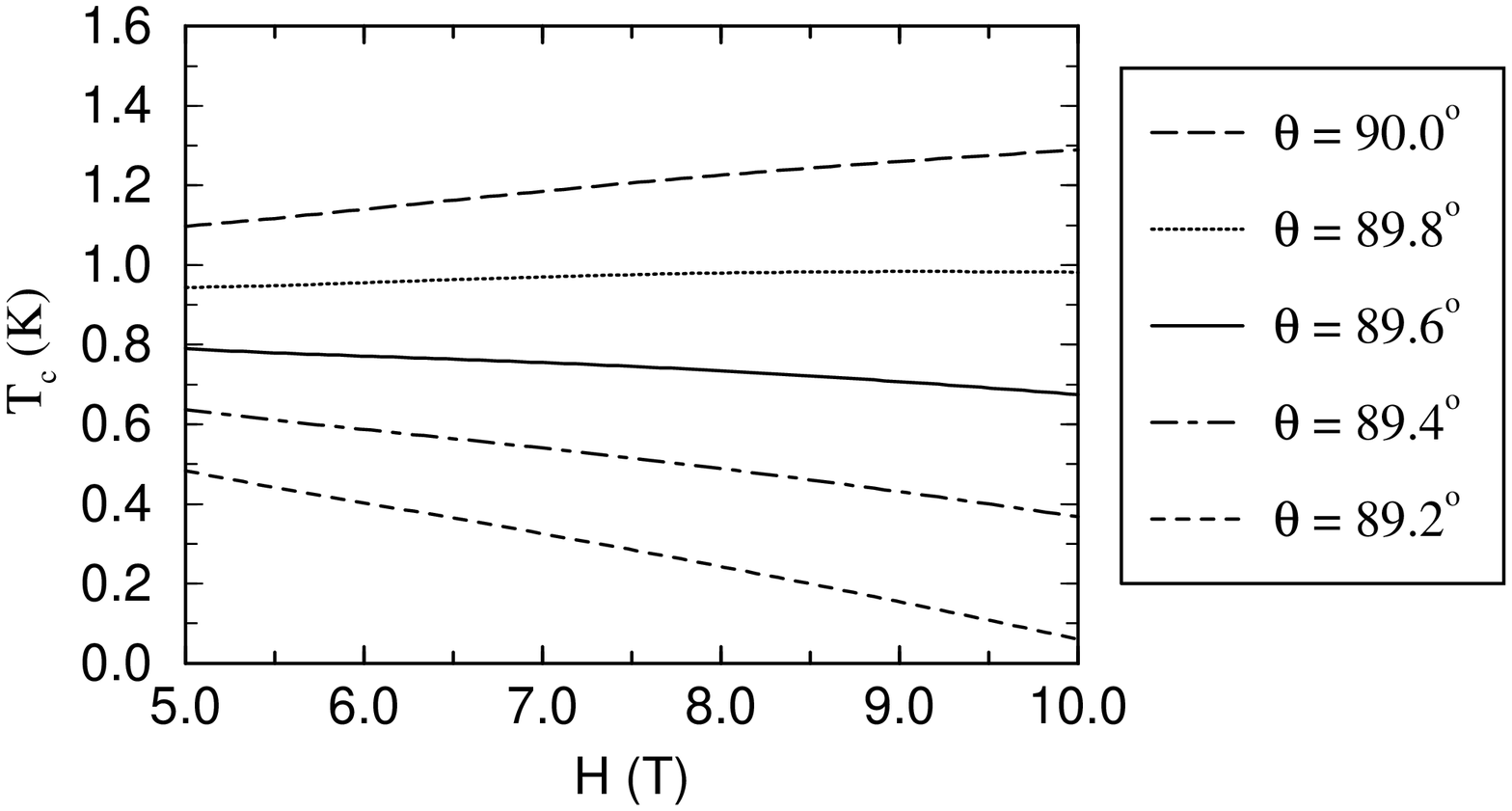}
\vskip 0.5cm
\caption{This figure shows that the suppression 
of the critical temperature $T_c$ for applied fields
$H$ along the $yz$ plane in the two-dimensional 
regime. The direction $\theta = 90.0^{\circ}$ corresponds to perfect 
alignment along the $z$ axis. The parameters used were 
$T_{c_{2d}} = 1.5 {\rm K}$, $t_y = 100{\rm K}$, $t_z = 10{\rm K}$, 
with lattice constants characteristic of Bechgaard salts.}
\label{fig:tc2d}
\end{center}
\end{figure}

{\it Two-dimensional regime:}
When the value of 
magnetic field is lowered and the angle $\theta$ is changed to 
satisfy conditions $t_y/\omega_{c_y} \gg 1$ 
and $t_z/\omega_{c_z} \ll 1$,
we reach a two dimensional regime where quantum confinement occurs only
along the $z$ direction. In this two dimensional regime the effects of 
the magnetic field along the y direction can be treated semiclassically.
Using Eq.~(\ref{eqn:deltaxyz}) the resulting eigenvalue equation is 
\begin{equation}
\label{eqn:gap2d}
\left[ 
-\beta_{2d}{\ell}_x^2 {d^2 \over dx^2} + L_y (x) 
+ A_{2d} + B_z \cos\phi_z(x) 
\right] u(x) = 0,
\end{equation}
where 
the coefficient of the first term is
$\beta_{2d} = ( 1 - t_z^2/\omega_{c_z}^2 )$.
The second term is
$L_y (x) = ({\ell}_y \phi_y (x) /b)^2$,
with coefficient
${\ell}_y = \sqrt{C} t_y b/T\sqrt{2}$,
while the third term is
$$
A{_2d} = 
\beta_{2d} \ln ( {\bar\Omega} )
+ \left( {t_z \over \omega_{c_z} } \right)^2
\ln \left\vert { 2 \omega_d \over \omega_{c_z} }\right\vert
- {1 \over \lambda {\cal N} }.
$$
The critical temperature resulting from Eq.~(\ref{eqn:gap2d}) is
\begin{equation}
\label{eqn:tc2d}
T_c = T_{c_{2d}} 
\left[
1 - 
{ C \sqrt{2} t_y \omega_{c_y} 
\over T_{c_{2d}}^2 }
- 
\left( 
{t_z \over \omega_{c_z} }  
\right)^2  
\ln 
\left\vert 
{ \gamma \omega_{c_z} \over \pi T_{c_{2d}} } 
\right\vert
\right],
\end{equation}
where 
$
T_{c_{2d}} =
(2 \omega_d \gamma/ \pi) \exp (- 1/ \lambda {\cal N})
$~\cite{dos}. 
A plot of $T_c$ as a function of magnetic field for various angles
$\theta$ close to $90.0^{\circ}$ is shown in Fig.~(\ref{fig:tc2d}). 
\vskip -3cm

Notice in Fig.~(\ref{fig:tc2d}), 
the rapid reduction in $T_c$ for very 
small deviations from $\theta = 90.0^{\circ}$ 
$({\bf H} \parallel {\bf \hat y})$.
A similar drop in $T_c$ has been seen experimentally in 
the putative critical temperature of 
${\rm (TMTSF)_2 ClO_4}$ at high magnetic fields
for small angular deviations from the 
$y$ axis ($b^{\prime}$ axis)~\cite{lee-98}.


{\it Three-dimensional regime:}
The three dimensional regime is characterized by the absence of 
quantum confinement, and occurs only at low magnetic fields
where $t_y/\omega_{c_y} \gg 1$ and $t_z/\omega_{c_z} \gg 1$.
In this case the gap equation reduces to 
\begin{equation}
\label{gap3d}
\left[
- \ell_x^2 {d^2 \over dx^2} + L_y (x)
+L_z (x) + A_{3d} 
\right] u (x) = 0
\end{equation}
where $\ell_x$
and $L_y (x)$ were already defined, and 
$L_z (x) = ({\ell}_z \phi_z (x)/c)^2 $,
with characteristic length scale along the $z$ direction
${\ell}_z = \sqrt{C} t_z c/T\sqrt{2}$. The coefficient
$A_{3d} = \left[
\ln \left( {\bar \Omega} \right) - 1/{\lambda {\cal N}}
\right].
$
This low magnetic field regime corresponds to the
anisotropic Ginzburg-Landau limit, and thus the critical temperature
is 
\begin{equation}
\label{eqn:tc3d}
T_c = T_c (0) 
\left[ 
1 - 
{ C \sqrt{2} \over T_c^2 (0)}
\sqrt{ t_y^2 \omega_{c_y}^2 + t_z^2 \omega_{c_z}^2 }
\right].
\end{equation}
The angular dependence of the upper critical field in the $xy$ plane 
(instead of $yz$ plane discussed here) was studied by
Huang and Maki~\cite{maki-89} in the Ginzburg-Landau regime. 


In summary, 
we have assumed an ESTP state as a plausible candidate for triplet
superconductivity in quasi-one-dimensional 
systems~\cite{sdm-99,sdm-96,sdm-98},
and we have presented analytical results for the angular dependence 
of the upper critical field (critical temperature) of 
these superconductors when the magnetic field is
applied in the $yz$ plane. We have discussed three general regimes:
(a) a one-dimensional regime at very high magnetic fields where there is 
quantum confinement both along the $y$ and $z$ directions 
($t_y /\omega_{c_y} \ll 1$ and $t_z /\omega_{c_z} \ll 1$); 
(b) a two-dimensional regime where there
is quantum confinement only along the $z$ direction 
($t_y /\omega_{c_y} \gg 1$ and $t_z /\omega_{c_z} \ll 1$); and 
(c) a three-dimensional regime at low magnetic fields 
(Ginzburg-Landau limit,
where there is no quantum confinement 
($t_y /\omega_{c_y} \gg 1$ and $t_z /\omega_{c_z} \gg 1$).
The one-dimensional limit is reached for quite high magnetic fields
$H \ge 100 {\rm T}$, which are not available today from continuous 
field sources. However, the two-dimensional regime, 
which is in the range of 
$5$ to $30 {\rm T}$, is attainable with continuous field
sources. This two-dimensional regime is very interesting 
from the experimental point of view, because there is a 
very rapid suppression of the predicted reentrant 
superconducting phase 
(for ${\bf H} \parallel {\bf \hat y}$)~\cite{lebed-86,lebed-87,dms-93},
when the applied magnetic field is just fractions of a
degree away from perfect alignment with the $y$ $(b^{\prime})$ axis.
Our analytical calculations seem to be in qualitative agreement with
the possible experimental observation of such an effect in 
${\rm (TMTSF)_2 ClO_4}$~\cite{lee-98}. We also have shown that the 
three dimensional regime just corresponds to the usual anisotropic
Ginzburg-Landau limit. 
We would like to thank the Georgia Institute of Technology, 
NSF (Grant No. DMR-9803111) and 
NATO (Grant No. CRG-972261) for financial support.


\begin{references}
\bibitem{chi-00} In the weak spin-orbit coupling 
ESTP state, the magnetic free energy 
$F_{mag} = - \chi_{\alpha \beta}H_{\alpha} H_{\beta}/2$ is 
minimized when the spin quantization axis (orientation
of the spins of the Cooper pairs) coincides with the direction of 
the applied magnetic field, i.e., when the maximal eigenvalue of
$\chi_{\alpha \beta}$ occurs along the direction of 
the applied magnetic field.
The maximal eigenvalue at zero temperature and low fields is
$\chi_n$ for any direction of the applied magnetic field.
Thus, until measurements of Knight shifts (spin susceptibilities) along
different directions is performed, the weak spin-orbit coupling
ESTP state remains a very promising candidate for the order parameter
symmetry of the ${\rm (TMTSF)_2 X}$ family.
\bibitem{triclinic-00} 
The Bechgaard salt family ${\rm (TMTSF)_2 X}$ is truly
triclinic, however, many of the qualitative results discussed in this
paper can be easily generalized to the triclinic case.
\bibitem{lambdamu} In general both $\lambda_{\mu}$ and ${\cal N}_{\mu}$
could have different values depending on the $m_s$ channel, however were
they different, there would be two distinct critical temperatures 
for $m_s = 1$ ($\mu = \uparrow\uparrow$) and 
$m_s = -1$ ($\mu = \downarrow\downarrow$), respectively.
\bibitem{dos} Here we have assumed that
both the interaction $\lambda$ and the single particle density
of states ${\cal N}$ are independent of the magnetic field, 
however, both $\lambda$ and ${\cal N}$ can be renormalized 
by the magnetic field, thus making $T_{c_{1d}}$ or $T_{c_{2d}}$ 
different from $T_c (0)$.
\end{references}
\end{document}